\documentclass[a4paper]{jpconf}
\usepackage{graphicx,amsmath,alltt}

\DeclareSymbolFont{ttoperators}{OT1}{cmtt}{m}{n}
\newcommand\xCode[1]{{%
  \mathcode`\"="0\the\symttoperators22%
  \mathchardef\$="4\the\symttoperators24%
  \mathcode`\(="4\the\symttoperators28%
  \mathcode`\)="5\the\symttoperators29%
  \mathcode`\/="0\the\symttoperators2F%
  \mathcode`\[="4\the\symttoperators5B%
  \mathcode`\]="5\the\symttoperators5D%
  \mathchardef\{="4\the\symttoperators7B%
  \mathchardef\}="5\the\symttoperators7D%
  \ensuremath{\mathtt{#1}}}}
\newcommand\Code[1]{\texttt{#1}}
\newcommand\eg{e.g.\ }
\newcommand\ie{i.e.\ }
\newcommand{\rd}{\mathrm{d}}
\newcommand\greyed[1]{\special{ps: .5 setgray}#1\special{ps: 0 setgray}}

\begin{document}

\title{FormCalc 7}

\author{S.~Agrawal$^1$, T.~Hahn$^2$, E.~Mirabella$^2$}

\address{%
${}^1$ LNM IIT, Rupa Ki Nagal, Post-Sumel, Via Jamdoli, Jaipur--302031, India, \\
${}^2$ MPI f\"ur Physik, F\"ohringer Ring 6, D--80805 Munich, Germany}

\ead{\{shivam,hahn,mirabell\}@mpp.mpg.de}

\begin{abstract}
We present additions and improvements in Version 7 of FormCalc, most 
notably analytic tensor reduction, choice of OPP methods, 
and MSSM initialization via FeynHiggs, as well as a parallelized Cuba 
library for numerical integration.

\hfill Report MPP-2011-143
\end{abstract}

\section{Introduction}

The Mathematica package FormCalc \cite{FormCalc} simplifies Feynman 
diagrams generated with FeynArts \cite{FeynArts} up to one-loop order.  
It provides both the analytical results and can generate Fortran code 
for the numerical evaluation of the squared matrix element.

This note presents the following improvements and additions in FormCalc 
7 and the numerical integration package Cuba \cite{Cuba}, which is 
included in FormCalc:
\begin{itemize}
\item Analytic tensor reduction,

\item Unitarity methods (OPP),

\item Improved code generation,

\item Command-line parameters for model initialization,

\item MSSM initialization via FeynHiggs,

\item Built-in parallelization in Cuba.
\end{itemize}

\section{Analytic tensor reduction}

Despite the hype that surrounds unitarity methods today, the 
Passarino--Veltman decomposition of tensor one-loop integrals 
\cite{PaVe} remains a valuable technique, also because it admits a fully 
analytic reduction.  The complete tensor reduction consists of two 
steps:
\begin{itemize}
\item The Lorentz-covariant decomposition of the tensors of the loop 
momentum appearing in the numerator into linear combinations of tensors 
constructed from $g_{\mu\nu}$ and the external momenta with coefficient 
functions, \eg
$$
\int\rd^4 q\frac{q_\mu q_\nu}{D_0 D_1} \sim
B_{\mu\nu} = g_{\mu\nu} B_{00} + p_\mu p_\nu B_{11}\,.
$$
This part has always been performed in FormCalc, as the actual tensors 
are rather unwieldy objects for further evaluation.

\item Solving the linear system that determines the coefficient 
functions, \ie expressing the coefficient functions through scalar 
integrals.
\end{itemize}

FormCalc has for long included the add-on \Code{FormCalc`btensor`} 
package which analytically reduces one- and two-point functions when 
loaded, but higher-point functions could be reduced only indirectly 
through FeynCalc \cite{FeynCalc}, \ie the user had to save the 
amplitudes with \Code{FeynCalcPut}, run FeynCalc in a different 
Mathematica session, and load the reduced expressions into FormCalc 
again with \Code{FeynCalcGet}.  This procedure was not only suboptimal 
in terms of user-friendliness but also did not take advantage of the 
field levels of FeynArts, \ie FeynCalc always operated on the fully 
inserted amplitudes rather than the (typically much fewer) Generic 
amplitudes.

The analytic tensor reduction is meanwhile properly available in 
FormCalc and can be turned on through the option
\begin{alltt}
   CalcFeynAmp[..., PaVeReduce \(\to\) True]
\end{alltt}
Our code implements the reduction formulas of Denner and Dittmaier 
\cite{TensRed}.  While these are fully worked out, it nevertheless took 
considerable effort to program them in FORM due to at first sight 
trivial issues, \eg that there is no straightforward way to obtain the 
$N$-th argument of a function.  Adding the reduction code to the 
Mathematica part of FormCalc instead was not an option, however, as we 
wanted to operate on the Generic amplitudes, before the substitution of 
the insertions, and this happens in FORM.

Inverse Gram determinants, which appear as a by-product of inverting the 
coefficient-function system, may lead to instabilities in the numerical 
evaluation later on and therefore FormCalc tries to cancel them as much 
as possible.  The ones that cannot be cancelled immediately are returned 
as \Code{IGram[$x$]} ($= 1/x$) and so can easily be found and processed 
further in Mathematica.

\section{Unitarity methods (OPP)}

We employ the OPP (Ossola, Papadopoulos, Pittau \cite{OPP}) unitarity 
method as implemented in the two libraries CutTools \cite{CutTools} and 
Samurai \cite{Samurai}.  Instead of introducing tensor coefficients as 
in the last section, in this case the whole numerator is placed in a 
subroutine, as in:
$$
\varepsilon_1^\mu\varepsilon_2^\nu B_{\mu\nu}(p, m_1^2, m_2^2) =
  B_{\mathrm{cut}}(2, N, p, m_1^2, m_2^2)\,,
\qquad\text{where}\qquad
N(q_\mu) = (\varepsilon_1\cdot q) \: (\varepsilon_2\cdot q)\,.
$$
The numerator subroutine $N$ will be sampled by the OPP function 
($B_{\mathrm{cut}}$ in this example).  The first argument of
$B_{\mathrm{cut}}$, 2, refers to the maximum power of $q$ in $N$.

Subexpressions of the numerator function (coefficients, summands, etc.) 
which do not depend on $q$ are pulled out and computed once, ahead of 
invoking the OPP function, using FormCalc's abbreviationing machinery 
\cite{FCabbr}.  In particular in BSM theories, these coefficients can be 
lengthy such that pulling them out significantly increases performance.

Our implementation admits mixing Passarino--Veltman decomposition with 
OPP in the sense that one chooses an integer $n$ starting from which an 
$n$-point function is treated with OPP methods.  For example, \Code{OPP 
$\to$ 4} means that $A$, $B$, $C$ functions are treated with 
Passarino--Veltman and $D$ and up with OPP.  A negative $n$ indicates 
that the rational terms for the OPP integrals shall be added 
analytically whereas else their computation is left to the OPP package.

The CutTools and Samurai libraries have minor differences in calling 
conventions but are otherwise similar enough to let the preprocessor 
handle the switching.  That is, one does not need to re-generate the 
Fortran code in order to link with the other library.  Specifically, the 
following steps must be taken in order to use the OPP method in 
FormCalc:
\begin{itemize}
\item The amplitudes must be prepared with \Code{CalcFeynAmp[..., OPP 
$\to n$]} ($n < \infty$).

\item In the generated code, the OPP library (CutTools or Samurai) must 
be chosen and the declarations in \Code{opp.h} be included.  This is 
most conveniently done in \Code{user.h}, in the following structure:
\begin{alltt}
\greyed{#ifndef USER_H}
\greyed{#define USER_H}
\greyed{* declarations for the whole file (e.g. preprocessor defs)}
#define SAMURAI                \greyed{\textrm{\emph{(or \texttt{CUTTOOLS})}}}
\greyed{#else}
\greyed{* declarations for every subroutine}
#include "opp.h"               \greyed{\textrm{\emph{(necessary for OPP)}}}
\greyed{#include "model_sm.h"}
\greyed{#endif}
\end{alltt}
\end{itemize}
As for testing, we have currently compared a handful of $2\to 2$ and 
$2\to 3$ scattering reactions, both QCD and electroweak, and found 
agreement to about 10 digits between Passarino--Veltman and OPP, with
CutTools and Samurai delivering similar results.  This shows that the 
method is working.

Performance is somewhat wanting as of now, however.  For the tested 
processes, quite frankly, Passarino--Veltman beats OPP hands down.  To 
be fair, OPP was in the first place designed to increase the reach of 
one-loop calculations to higher-leg multiplicities and not so much to 
speed up the ones with not so many legs such as those we have looked at 
so far.  Nevertheless, there seems to be room for improvement on a 
technical level and we are certainly continuing work on optimizing the 
performance.

\section{Improved code generation}

The main subroutine \Code{SquaredME.F} is now sectioned by comments, to 
aid automated substitution \eg with \Code{sed}.  For example, the 
variable declarations are enclosed in
\begin{verbatim}
   * BEGIN VARDECL
   ...
   * END VARDECL
\end{verbatim}
Furthermore, the generated code and the driver files are consistently 
formulated in terms of the newly introduced \Code{RealType} and 
\Code{ComplexType} data types, by default equivalent to \Code{double 
precision} and \Code{double complex}, respectively.  Note that 
capitalization matters as these words are substituted by the 
preprocessor.  This introduces a level of abstraction which makes it 
easier to \eg work with a different precision.

More importantly, work is ongoing to produce output in C (C99 to be 
precise, for support of complex numbers), in addition to the Fortran 
mode.  The advantages are threefold:
\begin{itemize}
\item It makes integration of generated code into existing C/C++ 
packages still easier (no linking hassles) while not impeding use in 
Fortran, \ie even though it is not planned to remove the native Fortran 
output in a future version of FormCalc, a Fortran user will be able to 
link with the C object files directly without any changes to his code.

\item It makes GPU programming feasible without reverting to commercial 
Fortran compilers.

\item One can take advantage of C's \Code{long double} data type which, 
at least on Intel x86 hardware, gives an additional 2--3 digits of 
precision at essentially no extra cost.  Extended real data types in 
Fortran, if available at all, are usually IEEE-754-compliant 
\Code{REAL*16} emulated in software.
\end{itemize}

\section{Command-line parameters for model initialization}

FormCalc includes a suite of so-called driver programs to manage the 
automatically generated code for computing the squared matrix element. 
They parse the command line, initialize model constants, set up phase 
space, etc.

In particular the driver modules for the initialization of the model 
parameters and luminosity calculation (which includes \eg the setup of 
PDFs used in hadronic reactions) had no access to the command-line 
arguments so far and could use only variables supplied by the user in 
the main control program \Code{run.F}.  In other words, the model inputs 
and PDF selections were `compiled in' and the executable had to be 
re-built every time those values changed.

The present command-line parser accepts so-called colon arguments 
(arguments starting with a `:') before the usual ones on the command 
line, as in:
\begin{alltt}
   run :\(\mathit{arg}\sb1\) :\(\mathit{arg}\sb2\) ... uuuuu 0,1000
\end{alltt}
The colon arguments are read into an array (sans colon) and handed to 
the model-initialization and luminosity-calculation subroutines:
\begin{verbatim}
   subroutine ModelDefaults(argc, argv)
\end{verbatim}
\begin{verbatim}
   subroutine LumiDefaults(argc, argv)
   integer argc
   character*128 argv(*)
\end{verbatim}
Note that, unlike in C (\Code{char **argv}), fixed-length strings are
passed in \Code{argv} since there are no pointers in Fortran 77.  It is 
up to the \Code{ModelDefaults} and \Code{LumiDefaults} subroutines to 
handle the arguments.  In Fortran it is furthermore no fatal error to
have no formal arguments in the \Code{ModelDefaults} and 
\Code{LumiDefaults} subroutines (as in previous FormCalc versions), so 
old code will compile and run without change.

\section{MSSM initialization via FeynHiggs}

The colon arguments of the previous section are immediately put to use 
for the initialization of the MSSM through FeynHiggs \cite{FeynHiggs}.
The default MSSM initialization is a stand-alone routine (\ie requires 
no external library to be linked), but is not quite as thorough as 
FeynHiggs when it comes to the corrections included \eg in the 
computation of the Higgs masses.

From FeynHiggs version 2.8.1 on not only the computational engine but 
the entire Frontend functionality is available through library routines 
so that the colon arguments can simply be passed to a FeynHiggs 
subroutine to make FeynHiggs initialize itself as if invoked from its 
own command-line Frontend.  The FormCalc-generated code inherits thus 
the ability to read parameter files in either native FeynHiggs or SLHA 
format, and of course obtains all MSSM parameters and Higgs observables 
from FeynHiggs. There is no duplication of initialization code this way, 
and moreover the parameters are consistent between the Higgs-mass and 
the cross- section calculations.

To use the FeynHiggs initialization, one chooses the 
model-initialization module \Code{model\_fh.F} instead of 
\Code{model\_mssm.F}.  The compiled code is invoked as
\begin{alltt}
   run :\(\mathit{parafile}\) \greyed{:\(\mathit{flags}\)} uuuuu 0,1000
\end{alltt}
The colon arguments are just the ones of the FeynHiggs Frontend: 
\textit{parafile} is the name of the parameter file and the optional 
\textit{flags} allows to override the default flags of FeynHiggs.

\section{Built-in parallelization in Cuba}

Cuba is a library for multidimensional numerical integration which is 
integrated in FormCalc but of course can be used independently, too.
Only the Mathematica interface was able to compute in parallel so far,
by redefining the function \Code{MapSample} with \eg \Code{ParallelMap}.
In the latest release we added parallelization also to the C/C++ and 
Fortran interfaces.

We attempt no parallelization across the network, say via MPI.  That is, 
we restrict ourselves to parallelization on one computer, using 
operating-system functions only, hence no extra software is required.  A 
common setup these days, even on laptops, is a single CPU with a number 
of cores, typically 4 or 8.  Utilizing many more compute nodes, as one 
could potentially do with MPI, is more of a theoretical option anyway 
since the speed-ups cannot be expected to grow linearly.

We use \Code{fork}/\Code{wait} rather than the \Code{pthread*} 
functions.  The latter are slightly more efficient at communicating data 
between parent and child because they share the same memory space (the 
\Code{fork}/\Code{wait} version uses pipes instead), but for the same 
reason they also require a reentrant integrand function, and apart from 
the extra work this takes, a programmer may not even have control over 
reentrancy in his language, \eg Fortran's I/O is typically 
non-reentrant.  \Code{fork} on the other hand creates a completely 
independent copy of the running process and thus works for any integrand 
function.

Changing the number of cores to use does not require a re-compile, which 
is particularly useful as the program image should be able to run on 
several computers (with possibly different numbers of cores) 
simultaneously.  Cuba determines the number of cores from the 
environment variable \Code{CUBACORES}, or if this is unset, takes the 
idle cores on the present system (total cores minus load average).  That 
is, unless the user explicitly sets \Code{CUBACORES}, a program calling 
a Cuba routine will automatically parallelize on the available cores.  A 
master process orchestrates the parallelization but does not count 
towards the number of cores, \eg \Code{CUBACORES = 4} means four workers 
and one master.  Very importantly, the samples are generated by the 
master process only and distributed to the workers, such that random 
numbers are never used more than once.

\begin{figure}
\includegraphics[width=\hsize]{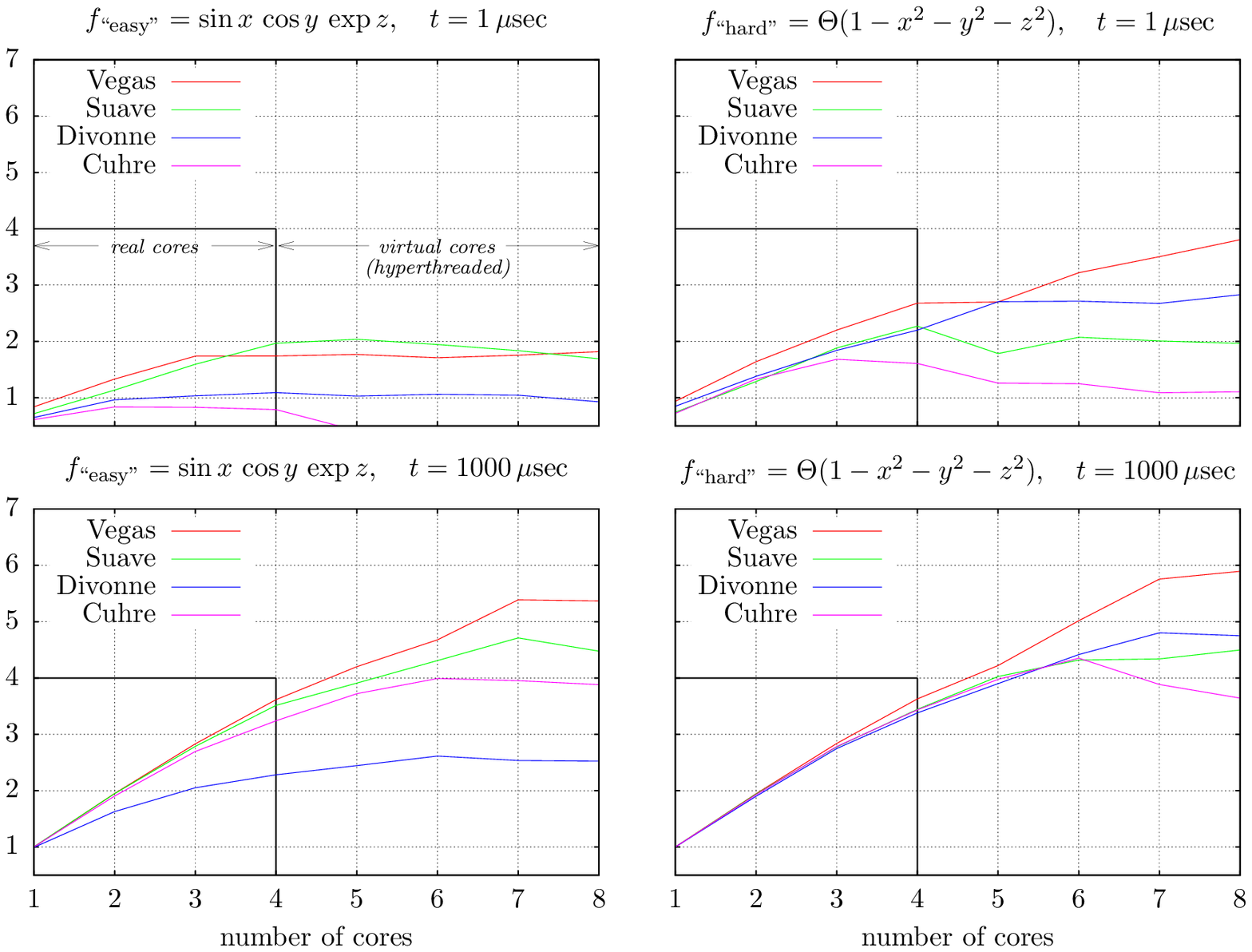}
\caption{\label{fig:speedups}Cuba speed-ups for a three-dimensional 
integral on an i7-2600 Linux system (2.6.37) with 4 real/8 virtual 
(hyperthreaded) cores.  The vertical line at 4 cores marks the 
cross-over. 

Left column: `easy' integrand, right column: `hard' integrand.

Top row: `fast' integrand ($1\,\mu$sec), bottom row: `slow' integrand 
($1000\,\mu$sec per evaluation).

Note that also in the one-core case a parallel version is used (one 
master, one worker), which explains why the timings normalized to the 
serial version are below 1, in the top row visibly so.

The measured speed-ups are, if anything, on the conservative side as the 
i7 CPUs boast a feature named Turbo Boost which allows the CPU to scale 
up the frequency if not all cores are loaded, \ie the serial version 
will likely have run at a somewhat higher CPU frequency.}
\end{figure}

Parallelization entails a certain overhead as usual, so the efficiency 
will depend on the `cost' of an integrand evaluation, \ie the more 
`expensive' (time-consuming) it is to sample the integrand, the better 
the speed-up will be.  To give an idea of the values that can be 
attained, Fig.\ \ref{fig:speedups} shows the speed-ups for an `easy' and 
a `hard' one of the 11 integrands of the demo program included in the 
Cuba package for two different integrand delays.  To tune the `cost' of 
the integrands, we introduced a calibrated delay loop into the integrand 
functions (which are simple one-liners and for our purposes `infinitely' 
fast).  The calibration and the timing measurements are rather delicate 
and shall not be discussed here.

The first, expected, observation is that parallelization is worthwhile 
only for not-too-fast integrands.  This is not a major showstopper, 
however, as many integrands in particle physics (one-loop 
cross-sections, for example) easily fall into the 
1000-$\mu$sec-and-beyond category.

The second observation is that parallelization works best for
`simple-minded' integrators, \eg Vegas.  The `intelligent' algorithms 
are generally much harder to parallelize because they don't just do 
mechanical sampling but take into account intermediate results, make 
extra checks on the integrand (\eg try to find extrema), etc.  This is 
particularly true for Divonne, where we spent significant time to lift 
the speed-up curve even this far above 1 and which is still clearly 
lagging in parallelization efficiency (see \eg the bottom-left plot of 
Fig.~\ref{fig:speedups}).  Then again, the `intelligent' algorithms are 
usually faster to start with (\ie converge with fewer points sampled),
which compensates for the lack of parallelizability.

\section{Summary}

FormCalc 7 (\Code{http://feynarts.de/formcalc}) has many new and 
improved features, most notably the analytic tensor reduction, choice of 
OPP methods, and the link with FeynHiggs.  The Cuba library 
(\Code{http://feynarts.de/cuba}), also included in FormCalc, 
parallelizes integrations automatically and achieves decent speed-ups 
for typical cross-section integrands.

\pagebreak

\medskip
\raggedright

\end{document}